\documentclass[twoside,twocolumn,9pt]{article}
\usepackage{extsizes}
\usepackage[super,sort&compress,comma]{natbib}
\usepackage[version=3]{mhchem}
\usepackage[left=1.5cm, right=1.5cm, top=1.785cm, bottom=2.0cm]{geometry}
\usepackage{balance}
\usepackage{mathptmx}
\usepackage{sectsty}
\usepackage{graphicx}
\usepackage{lastpage}
\usepackage[format=plain,justification=justified,singlelinecheck=false,font={stretch=1.125,small,sf},labelfont=bf,labelsep=space]{caption}
\usepackage{float}
\usepackage{fancyhdr}
\usepackage{fnpos}
\usepackage[english]{babel}
\addto{\captionsenglish}{%
  
}
\usepackage{array}
\usepackage{droidsans}
\usepackage{charter}
\usepackage[T1]{fontenc}
\usepackage[usenames,dvipsnames]{xcolor}
\usepackage{setspace}
\usepackage[compact]{titlesec}
\usepackage{hyperref}

\usepackage{physics}
\usepackage{wasysym}
\usepackage{xcolor}
\usepackage{amsmath}
\usepackage{amssymb}

\definecolor{cream}{RGB}{222,217,201}

\begin{document}

\pagestyle{fancy}
\thispagestyle{plain}
\fancypagestyle{plain}{
    \renewcommand{\headrulewidth}{0pt}
}

\makeFNbottom
\makeatletter
\renewcommand\LARGE{\@setfontsize\LARGE{15pt}{17}}
\renewcommand\Large{\@setfontsize\Large{12pt}{14}}
\renewcommand\large{\@setfontsize\large{10pt}{12}}
\renewcommand\footnotesize{\@setfontsize\footnotesize{7pt}{10}}
\makeatother

\renewcommand{\thefootnote}{\fnsymbol{footnote}}
\renewcommand\footnoterule{\vspace*{1pt}%
    \color{cream}\hrule width 3.5in height 0.4pt \color{black}\vspace*{5pt}}
\setcounter{secnumdepth}{5}

\makeatletter
\renewcommand\@biblabel[1]{#1}
\renewcommand\@makefntext[1]%
{\noindent\makebox[0pt][r]{\@thefnmark\,}#1}
\makeatother
\renewcommand{\figurename}{\small{Fig.}~}
\sectionfont{\sffamily\Large}
\subsectionfont{\normalsize}
\subsubsectionfont{\bf}
\setstretch{1.125} 
\setlength{\skip\footins}{0.8cm}
\setlength{\footnotesep}{0.25cm}
\setlength{\jot}{10pt}
\titlespacing*{\section}{0pt}{4pt}{4pt}
\titlespacing*{\subsection}{0pt}{15pt}{1pt}
\fancyfoot{}
\fancyfoot[LO,RE]{\vspace{-7.1pt}\includegraphics[height=9pt]{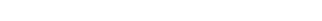}}
\fancyfoot[CO]{\vspace{-7.1pt}\hspace{13.2cm}\includegraphics{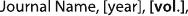}}
\fancyfoot[CE]{\vspace{-7.2pt}\hspace{-14.2cm}\includegraphics{head_foot/RF}}
\fancyfoot[RO]{\footnotesize{\sffamily{1--\pageref{LastPage} ~\textbar  \hspace{2pt}\thepage}}}
\fancyfoot[LE]{\footnotesize{\sffamily{\thepage~\textbar\hspace{3.45cm} 1--\pageref{LastPage}}}}
\fancyhead{}
\renewcommand{\headrulewidth}{0pt}
\renewcommand{\footrulewidth}{0pt}
\setlength{\arrayrulewidth}{1pt}
\setlength{\columnsep}{6.5mm}
\setlength\bibsep{1pt}

\makeatletter
\newlength{\figrulesep}
\setlength{\figrulesep}{0.5\textfloatsep}

\newcommand{\topfigrule}{\vspace*{-1pt}%
    \noindent{\color{cream}\rule[-\figrulesep]{\columnwidth}{1.5pt}} }

\newcommand{\botfigrule}{\vspace*{-2pt}%
    \noindent{\color{cream}\rule[\figrulesep]{\columnwidth}{1.5pt}} }

\newcommand{\dblfigrule}{\vspace*{-1pt}%
    \noindent{\color{cream}\rule[-\figrulesep]{\textwidth}{1.5pt}} }
 
\makeatother


\twocolumn[
    \begin{@twocolumnfalse}
        \vspace{1em}
        \sffamily
        \begin{tabular}{m{4.5cm} p{13.5cm} }
            \includegraphics{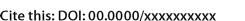}
            & \noindent\LARGE{\textbf{Chiral fluid membranes with orientational order and multiple edges}}\\
            \vspace{0.3cm} & \vspace{0.3cm}\\
            & \noindent\large{Lijie Ding,$^{\ast}$\textit{$^{a}$} Robert A. Pelcovits,\textit{$^{ab}$} and Thomas R. Powers\textit{$^{abcd}$}}\\
            \includegraphics{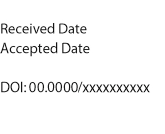} & \noindent\normalsize{We carry out Monte Carlo simulations on fluid membranes with orientational order and multiple edges in the presence and absence of external forces. The membrane resists bending and has an edge tension, the orientational order couples with the membrane surface normal through a cost for tilting, and there is a chiral liquid crystalline interaction. In the absence of external forces, a membrane initialized as a vesicle will form a disk at low chirality, with the directors forming a smectic-A phase with alignment perpendicular to the membrane surface except near the edge. At large chirality a catenoid-like shape or a trinoid-like shape is formed, depending on the number of edges in the initial vesicle. This shape change is accompanied by cholesteric ordering of the directors and multiple $\pi$ walls connecting the membrane edges and wrapping around the membrane neck. If the membrane is initialized instead in a cylindrical shape and stretched by an external force, it maintains a nearly cylindrical shape but additional liquid crystalline phases appear. For large tilt coupling and low chirality, a smectic-A phase forms. For lower values of the tilt coupling, a nematic phase appears at zero chirality with the average director oriented perpendicular to the long axis of the membrane, while for nonzero chirality a cholesteric phase appears. The $\pi$ walls are tilt walls at low chirality and transition to twist walls as chirality is increased. We construct a continuum model of the director field to explain this behavior. 
            }\\
        \end{tabular}
    \end{@twocolumnfalse} \vspace{0.6cm}
]

\renewcommand*\rmdefault{bch}\normalfont\upshape
\rmfamily
\section*{}
\vspace{-1cm}

\footnotetext{\textit{$^{\ast}$~Email: Lijie\_Ding@alumni.brown.edu}}
\footnotetext{\textit{$^{a}$~Department of Physics, Brown University, Providence, RI 02912, USA.}}
\footnotetext{\textit{$^{b}$~Brown Theoretical Physics Center, Brown University, Providence, RI 02912, USA.}}
\footnotetext{\textit{$^{c}$~School of Engineering, Brown University, Providence, RI 02912, USA.}}
\footnotetext{\textit{$^{d}$~Center for Fluid Mechanics, Brown University, Providence, RI 02912, USA.}}




\section{Introduction}
\label{sec:introduction}
    Many structures formed by fluid membranes or thin films with liquid crystalline degrees of freedom result from the interplay of curvature and orientational order. For example, the ordering of curved rod-like proteins in cell membranes can lead to the formation of cylindrical membrane shapes.~\cite{FrostUngerCamilli2009} Tubules can also be formed in lipid membranes due to the chirality of the lipid molecules.~\cite{Georger_etal1987,SelingerSchnur1993} Liquid crystalline shells~\cite{Lopez-LeonFernandez-Nieves2011} provide another example, with nematic and cholesteric~\cite{Tran_etal2017,Carenza_etal2022} textures arising from the interaction of the curvature of the shell and the liquid crystalline order.~\cite{napoli2021nematic,nitschke2020liquid,napoli2012extrinsic} Other recent examples include colloidal membranes made of chiral filaments, such as rod-like fd viruses~\cite{gibaud2017filamentous} or DNA origami filaments.~\cite{siavashpouri2017molecular} A key distinction between the colloidal membranes and the other examples is that colloidal membranes tend to have free edges, which is the focus of this work.

    Colloidal membranes are single layer liquid crystal structures of filaments assembled through a depletion force. These filaments form a cholesteric phase when concentrated in bulk\cite{barry2009model,siavashpouri2017molecular} which indicates they have chirality and tend to twist about each other. Changing the concentration of the depletant and the temperature leads to various structures of colloidal membranes,\cite{gibaud2017filamentous,saddles} including tactoids, disks, twisted ribbons, stacked membranes, saddles, catenoids, trinoids, four-noids, and higher order structures. When single layer colloidal membranes are formed, the filaments that comprise the membrane tend to align with the membrane surface normal, and the twist of the filaments is expelled to the edge of the membrane.\cite{gibaud2012reconfigurable} However, flat colloidal membranes can also sustain significant twist at interior points. For example, coalescence of two disk-shaped membranes\cite{zakhary2014imprintable} can lead to the formation of $\pi$-walls where the filaments rotate through $180^\circ$, making an angle of $90^\circ$ with the surface normal at the midpoint of the wall.

    Theoretical studies of the role of chirality in membrane shape have long been of interest for membranes composed of chiral lipid molecules.\cite{Selinger2001} Helfrich and Prost\cite{HelfrichProst} introduced a term linking molecular chirality to membrane bending, shown later to be identical (up to a line integral) to the Frank elastic term linear in director twist on a curved surface.\cite{OuYangLiu} Selinger et al\cite{SelingerSchnur1993,selinger1996theory,Selinger2001} studied tubules with helically modulated tilting states and helical ripples. Tu and Seifert\cite{TuSeifert2007} considered a concise theory of chiral membranes, deriving Euler-Lagrange equations assuming constant tilt of the molecules relative to the layer normal. Their model was extended by Kaplan et al.\cite{kaplan2010theory} to include variations in the tilt angle. All of these studies required an \textit{a priori} assumption of the shape of the membrane, e.g., tubules (with uniform tilt or a helically modulated tilt state), helical stripes, or twisted ribbons.

    Previously, \cite{ding2020shapes,ding2021deformation} we developed a Monte Carlo simulation scheme that allows for arbitrary membrane shapes using a discretized effective energy based on the continuum energy used by Kaplan et al.\cite{kaplan2010theory} The membrane surface was modeled by a triangular mesh with beads on the vertices connected by bonds, and the orientational order was modeled by unit vectors decorating each bead on the mesh. The energy for the discrete membrane included both the energy for the membrane shape and the liquid-crystalline energy for the orientational order. We used the Canham-Helfrich bending energy with an edge tension for the energy of membrane shape, a chiral Lebwohl-Lasher model for the director-director interaction, and a tilt coupling for the interaction between directors and membrane surface normal. We found the formation of a cholesteric phase at large chirality with the development of ripples in the surface due to the coupling between surface shape and orientational order. Although our simulations accounted for both the deformation of the membrane surface and the full orientational order of the constituent particles, they were limited to the simulation of single-edge membranes

    In this paper, we extend our previous study of the chiral membranes with one edge and explore the structure of chiral membranes with multiple edges. We start by investigating the role of chirality in the equilibrium shapes of multi-edge membranes. When initialized as a vesicle, a membrane with two edges can form a catenoid-like shape with cholesteric order if the chirality is sufficiently large. A membrane with three edges can undergo an additional transition at a higher value of chirality to a trinoid-like shape again with cholesteric order. Next, we apply a force to the edges of a membrane initialized in a cylindrical shape and find that stretching the membrane leads to the appearance of nematic and smectic-A phases in addition to cholesteric as the membrane adopts a nearly cylindrical shape. In the cholesteric phase of both the unstretched catenoid and cylinder, $\pi$-walls appear in the director field joining the two edges. At low chirality, the walls are tilt walls, while at higher chirality they are twist walls. We present a continuum analytical model that shows how the structure of the $\pi$-walls is determined by the liquid crystalline parameters.
    
\section{Model and method}
\label{sec:model_and_method}
    As in our previous work,\cite{ding2020shapes, ding2021deformation} we model the membrane using a dynamical beads-and-bonds triangular mesh $\mathcal{M},$\cite{gompper1997network} with hard beads of diameter $\sigma_0$ located at each vertex of the triangular mesh. The beads are connected by bonds of maximum length $l_0$. Each vertex $i$ of the mesh has a unit length director field $\vu{u}_i$. 
    
    The total energy $E$ of the membrane is the sum of a surface energy $E_{s}$ that depends on the geometric properties of the triangular mesh and a liquid-crystalline energy $E_{lc}$ that depends on the director field and its coupling to the mesh. The surface energy $E_{s}$ has contributions arising from the discretized Canham-Helfrich bending energy\cite{ding2020shapes,canham1970minimum,helfrich1973elastic} and a line tension of the edge:
    \begin{equation}
        E_s = \frac{\kappa}{2}\sum_{i\in\mathring{\mathcal{M}}} (2H_i)^2\sigma_i +\lambda \sum_{i\in \partial \mathcal{M}}\dd{s}_i,
        \label{eq:E_s}
    \end{equation}
    where $\kappa$ is the bending modulus, and $H_i$ and $\sigma_i$ are the mean curvature and the area of the cell on the virtual dual lattice at bead $i$, respectively. Complete expressions for each of the terms above can be found elsewhere.\cite{espriu1987triangulated,gompper1997network,ding2020shapes} In the last term above, $\lambda$ is the line tension of the edge and $\dd{s}_i$ is the differential edge length at bead $i$. The first summation in eqn (\ref{eq:E_s}) is over all interior beads $\mathring{\mathcal{M}}$ of the mesh and the second summation is over all beads $\partial\mathcal{M}$ on the edges.

    The liquid-crystalline energy $E_{lc}$ consists of three contributions: the director-director coupling, the chiral energy and the director-surface coupling:
    \begin{equation}
        \begin{aligned}
            E_{lc} =& -\epsilon_{LL} \sum_{(i,j)\in\mathcal{B}} \left[\frac{3}{2}(\vu{u}_i\cdot\vu{u}_j)^2-\frac{1}{2}\right] \\
            & -\epsilon_{LL} k_c \sum_{(i,j)\in\mathcal{B}}(\vu{u}_i\cross\vu{u}_j)\cdot\vu{r}_{ij}(\vu{u}_i\cdot\vu{u}_j) \\
            & + \frac{1}{2}C\sum_{i\in\mathcal{M}} \left[1-(\vu{u}_i\cdot \vu{n}_i)^2\right].
        \end{aligned}
        \label{eq:E_lc}
    \end{equation}
    The first term on the right-hand side of eqn (\ref{eq:E_lc}) is the Lebwohl-Lasher interaction\cite{lebwohl1972nematic}, which favors the alignment of neighboring directors on the triangular mesh. The second term is the chiral Lebwohl-Lasher interaction,\cite{vanderMeer1976} which favors a right-handed twist between neighboring directors when $k_c>0$. The separation $\hat r_{ij}$ is the direction from vertex $i$ to vertex $j$, and the product $(\vu{u}_i\cdot\vu{u}_j)$ is included to satisfy the plus-minus symmetry of the director field. The final term represents the tilt coupling of the director field to the local surface normal $\vu{n}_i$ of the triangular mesh at bead $i$, favoring alignment of the director and the surface normal, a tendency \cite{barry2009direct} arising from the depletion interaction. \cite{1954AO} A detailed expression for the surface normal can be found elsewhere.\cite{ding2020shapes,Crane:2013:DGP} The summations in the Lebwohl-Lasher and chiral interactions are over all bonds $\mathcal{B}$ in the mesh and the summation in the tilt energy is over all beads $\mathcal{M}$, both in the interior and on the edges.

\begin{figure}
        \centering
        \includegraphics[width=1\linewidth]{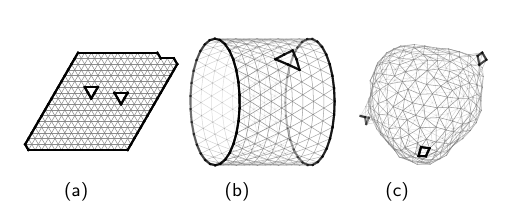}
        \caption[Initial configuration of a triangular mesh with three edges]{Possible initial configurations of a triangular mesh with three edges and $N=300$ beads. The internal bonds on the mesh are represented by thin gray segments and the edge bonds are represented by thick black lines. (a) A flat membrane with two holes. (b) A cylindrical membrane with one hole. (c) A vesicle with three holes.}
        \label{fig:init_config_demo}
    \end{figure}

    Monte Carlo simulations are carried out using this model by updating both the beads-and-bonds triangular mesh and the director field. Both the shape update and director update follow the same procedure as described in our previous papers.\cite{ding2020shapes,ding2021deformation} The new element in this paper is that the membranes have more than one edge, meaning that there can be holes in the membrane. Fig.~\ref{fig:init_config_demo} shows possible initial configurations of a triangular mesh with three edges. In Fig.~\ref{fig:init_config_demo}a, a flat membrane has two small triangular holes representing the second and third edges, and the cylinder in Fig.~\ref{fig:init_config_demo}b has one triangular hole for the third edge. A third possibility is a vesicle with three holes on the mesh as shown in Fig.~\ref{fig:init_config_demo}c. In principle these holes should disappear by nucleation. However, because we do not have an edge creation or edge removal update, we insert the holes by hand in order to create multiple edges.

    For a system of $N$ beads, each Monte Carlo (MC) step is composed of $N/t^2$ attempts to move a bead chosen at random, $2N/t^2$ attempts to flip a bond chosen at random and $\sqrt{N}/t^2$ attempts to shrink or extend the edge of the membrane. The parameter $t$ is defined by the bead move update which randomly selects one bead from all the beads on the mesh with equal probability and moves it to a random position in a cube of side $2t$ centered at its current position. The bond flip update selects a bond from all of the bonds in the interior of the triangular mesh with equal probability, detaches it from the beads at its endpoints and then flips it to connect the two opposite beads on the adjacent triangle. The parameter $t$ is set to $0.1$, with all lengths measured in units of the bead diameter $\sigma_0$. Energies are measured in units of $k_B T$.  The maximum length of the bonds on the triangular mesh is set to be $l_0=1.73<\sqrt{3}$ to satisfy self-avoidance and ensure fluidity of the membrane.\cite{gompper2000melting} During the simulation, $1.7\times 10^4$ MC steps were performed. To help the system reach equilibrium, we first carry out a simulated annealing for $2\times 10^3$ MC steps starting at infinite temperature where  $\beta\equiv1/k_BT=0$ and lower the temperature by  increasing $\beta$ in steps of $\delta\beta=0.01$ until $\beta=1$. We then equilibrate the system for another $5\times 10^3$ MC steps, and record observables for the remaining $1\times 10^4$ MC steps.

\section{Equilibrium shapes of membranes with multiple edges}
\label{sec:Equilibrium_shapes_of_membranes_with_multiple_edges}
\begin{figure}[thb!]
        \includegraphics[width=\linewidth]{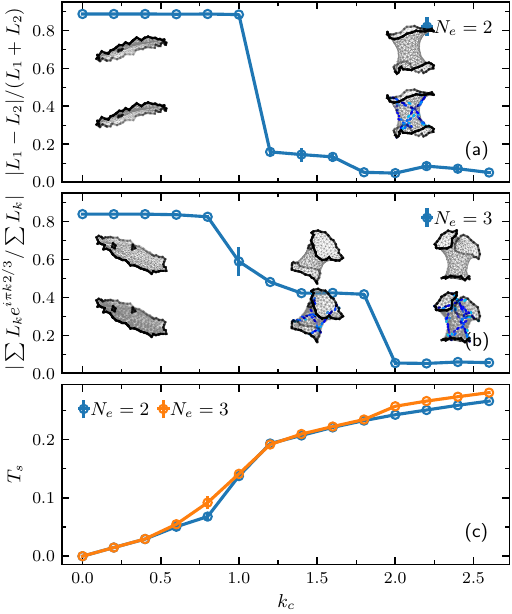}
        \caption{Shapes and features of three-dimensional membrane structures for varying chirality $k_c$. The initial configuration for each value of chirality is a vesicle with total number of edges $N_e = 2$ and $N_e=3$ for (a) and (b) respectively. The edges bound small holes of size comparable to a triangle on the  mesh [see Fig.~\ref{fig:init_config_demo}(c) for an example with $N_e=3$].  Here, the number of beads $N=300$, the bending modulus $\kappa=30$, the edge tension $\lambda=6$, and the Lebwohl-Lasher constant and tilt coupling $\epsilon_{LL}=C=4$.
        (a) Asymmetry of the length of the edges versus chirality for a two-edge membrane. The membrane shapes in the top row show the triangular mesh without the directors. The membrane shapes in the bottom row show the directors which lie on the $\pi$ walls that wind around the membrane and join the edges.
        (b) Same as (a) except for a membrane with three edges labeled with $k=0,1,2$. 
        (c) Average twist of the director field $T_s=\left<(\vu{u}_i\cross\vu{u}_j)\cdot \vu{r}_{ij} (\vu{u}_i\cdot\vu{u}_j)\right>_{(i,j)}$ as a function of chirality for the membranes shown in (a) and (b), where the average $\left<\dots\right>_{(i,j)}$ is over all bonds $(i,j)$.}
        \label{fig:topo_chirality}
    \end{figure}

    In our earlier work,\cite{ding2021deformation} we demonstrated that increasing the magnitude of the chirality of the director field leads to a rippling of a single-edge membrane coinciding with the appearance of cholesteric order. Here we study whether a potentially similar effect occurs in a multi-edge membrane. Fig.~\ref{fig:topo_chirality} shows that a membrane with two or three edges can transform respectively into shapes reminiscent of a catenoid or trinoid as chirality is increased. Below we will show quantitatively that these shapes approximate catenoids and trinoids, and therefore we will refer to them as such from here on. The drastic shape change of the membrane during these transformations involves crossing a very large free energy barrier and leads to strong hysteresis. We found that to consistently access these shapes, we must initialize the membrane as a vesicle with as many holes as edges, as in Fig.~\ref{fig:init_config_demo}c for the case of three edges. Initializing the shape in one of the other configurations shown in Fig.~\ref{fig:init_config_demo} can lead to very ``noisy'' shapes due to the large free energy barrier. The topology of the triangular mesh is fixed by the number of holes inserted into the vesicle due to the lack of an edge creation or removal update in our MC algorithm. Thus, even if one of the edges shrinks and its size becomes comparable to the triangles on the mesh, there will always be a small hole present.
    
    For a membrane with two edges (Fig.~\ref{fig:topo_chirality}a), one of the edges shrinks to a small hole when the chirality $k_c$ is small, and the membrane assumes a disk shape. The directors exhibit smectic-A order, i.e., they are aligned normal to the membrane except near the edges where they twist due to chirality.\cite{DEGENNEScholesteric} In our previous work \cite{ding2021deformation} where we did not insert a hole into the interior of the membrane, we found that as we increased chirality the membrane deformed into a saddle shape accompanied by the appearance of a cholesteric phase once $k_c \gtrsim 1$ . With the insertion of a hole in the initial membrane, a cholesteric phase again forms at a similar value of $k_c$, but accompanied by a much more dramatic shape change, namely, from a disk to a catenoid. The vertical axis in Fig.~\ref{fig:topo_chirality}a is the asymmetry of the length of the two edges, which is approximately $1$ when one edge dominates and approximately zero when both edges have nearly the same length.

    Fig.~\ref{fig:topo_chirality}b shows the result for a membrane with three edges. When the chirality is small, the membrane becomes a disk shape with two small holes. At intermediate values of chirality the membrane becomes a catenoid with one small hole in its interior. Further increasing chirality leads to an expansion of the small hole and the membrane become a trinoid shape with three edges of similar length. Such changes of shape are captured by the asymmetry of the length of the three edges shown on the vertical axis of the figure. When the membrane becomes a disk shape, one of the edges has a length much greater than the other two, i.e., $L_0\gg L_1\simeq L_2$. Thus, the measure of asymmetry becomes $|\sum L_k e^{i\pi k 2/3}/\sum L_k|\simeq |L_0/L_0|=1$. When the membrane forms a catenoid, we have $L_0\simeq L_1\gg L_2$, and the asymmetry becomes $|\sum L_k e^{i\pi k 2/3}/\sum L_k|\simeq |[L_0+L_1\cos(2\pi/3)+L_1i\sin(2\pi/3)]/(L_0+L_1)|\simeq 1/2$. Finally, when all three edges grow to have roughly the same perimeter, the asymmetry become approximately $0$.

    Fig.~\ref{fig:topo_chirality}c shows the average twist of the director field $T_s=\left<(\vu{u}_i\cross\vu{u}_j)\cdot \vu{r}_{ij} (\vu{u}_i\cdot\vu{u}_j)\right>_{(i,j)}$ for the membranes with two and three edges as a function of the chirality $k_c$, where the average $\left<\dots\right>_{(i,j)}$ is over all bonds $(i,j)$. The twist is positive, indicating a right-handed twist of the directors. Also, the $\pi$ walls on the catenoid and trinoid wind around the membrane in a right-handed sense. There is little difference between the average twist of the directors on the two structures, implying that the geometry of the $\pi$ walls is very similar on the two membranes.

    \begin{figure}[thb!]
        \includegraphics[width=\linewidth]{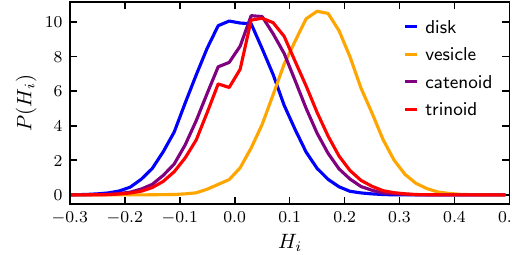}
        \caption{Probability distribution of the mean curvature $H_i$ (see eqn~(\ref{eq:E_s})) sampled over 200 MC configurations (50 MC steps between consecutive samples) for four membrane shapes: disk, vesicle, catenoid and trinoid. The membrane parameters are the same as in Fig.~\ref{fig:topo_chirality}, except for the vesicle where  $\lambda=50$. The chirality $k_c$ is zero for the disk and vesicle and 1.3 and 2.3 for the catenoid and trinoid, respectively.}
        \label{fig:mean_curvature}
    \end{figure}

    Fig.~\ref{fig:mean_curvature} shows a plot of the probability distribution of the mean curvature $H_i$ for the disk, catenoid and trinoid shapes shown in Fig.~\ref{fig:topo_chirality} and the vesicle shape shown in Fig.~\ref{fig:init_config_demo}c. The widths of the distributions for all four shapes are comparable and of the order of magnitude to be expected for thermal fluctuations where 
    \begin{equation}
      \sqrt{\langle H_i^2\rangle}\simeq \sqrt{\frac{k_B T}{\kappa\sigma_i}} \simeq 0.2,
    \end{equation}
    with $k_B T = 1$ (the final temperature of our simulations), $\kappa=30$ and $\sigma_i \simeq1$, where $\sigma_i$ is the area of the cell on the virtual dual lattice at bead $i$ (see eqn (\ref{eq:E_s})). The peaks of the distributions for the catenoid and trinoid are located at approximately equal values of $H_i$ which are substantially smaller than the location of the peak of the vesicle. While not mathematically minimal surfaces, the catenoids and trinoids appearing in our simulations are good approximations to ideal minimal surfaces.
  
    \begin{figure}[thb!]
        \centering
        \includegraphics[width=1\linewidth]{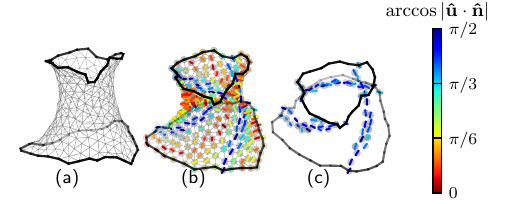}
        \caption{A sample configuration of a catenoid membrane. 
        (a) Triangular mesh model of the membrane shape. 
        (b) Hard beads decorated by directors $\vu{u}$ on each vertex are shown in addition to the triangular mesh, with the color indicating the angle $\arccos|\vu{u}\cdot\vu{n}|$ between the director and the local surface normal $\vu{n}$. 
        (c) Membrane edges and directors with $\arccos{|\vu{u}\cdot\vu{n}|}>\pi/3$, which are the directors approximately located in the $\pi$-walls. There are three such $\pi$-walls wrapping around the membrane and they begin and end on the membrane edges. For the sake of clarity the mesh and other directors are not shown in part (c).
        }
        \label{fig:config_demo}
    \end{figure}

    Fig.~\ref{fig:config_demo} shows a sample configuration of a catenoid including the directors. We note that there are three $\pi$-walls wrapping around the membrane, joining the two edges. The $\pi$-walls are characterized by a rotation of the director by $180^\circ$ about an axis which is perpendicular to the wall and lying in the local tangent plane of the membrane, as expected in the presence of cholesteric order. We saw similar lines in our previous study of the saddle shapes. The force-free catenoid shape as well as the tubule shapes we find when the catenoid is subject to external force (see next section) appear to be achiral (Fig.~\ref{fig:config_demo}a and ~\ref{fig:elongation_config_demo}), in contrast to the helical ribbons and tubules studied by Selinger et al.\cite{selinger1996theory} 

\section{Elongated membranes}
\label{sec:Elongated_membranes}
\subsection{Director field}
\label{ssec:director_field}
\begin{figure}[thb!]
        \centering
        \includegraphics[width=\linewidth]{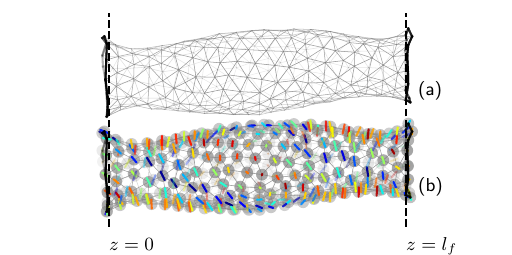}
        \caption{A sample configuration of an elongated membrane produced by applying equal and opposite forces to the edges of a cylindrical membrane. We need to state the values of $\kappa$, $C$, $\epsilon_{LL}$, and $k_c$. The edge beads are bounded by the condition $z\leq 0$ for the left edge and $z\geq l_f$ for the right edge, and black dashed lines mark the position of $z=0$ and $z=l_f$. 
        (a) Triangular mesh model of the membrane shape. 
        (b) Same as (a), but now including the directors attached to the beads. The color coding of the director orientation relative to the surface normal is the same as in Fig.~\ref{fig:config_demo}.}
        \label{fig:elongation_config_demo}
    \end{figure}

    We now focus on a two-edge membrane and study its response to an external force applied to the edges of a membrane initialized not as a vesicle as in the previous section, but rather in a cylindrical shape (similar to Fig.~\ref{fig:init_config_demo}b but without the small hole on the surface). Initializing in a cylindrical shape allows us to easily apply equal and opposite forces to the two edges which would be very difficult to accomplish for the catenoids shown in Fig.~\ref{fig:config_demo} where the edges are not planar. The force
    changes the length of the membrane and preempts the formation of a catenoid at large chirality, maintaining instead a nearly cylindrical shape except near the edges. The force is incorporated into our simulation by demanding that the beads on the right edge have $z\geq l_f$, and the beads on the left have $z\leq0$, where the $z$ axis is along the long axis of the cylinder and $l_f$ is the elongated length of the membrane under force. A sample configuration of an elongated membrane is shown in Fig.~\ref{fig:elongation_config_demo}. We specify $l_f$ in terms of inequalities on the $z$ coordinates of the edge beads to allow beads to join or leave the edges in our simulations. Constraining the edge beads to be exactly at $z=l_f$ or $z=0$ would prevent a bead in the interior from joining the edge unless it moves exactly to $z=l_f$ or $z=0$. By allowing the edge beads to move slightly into the $z\geq l_f$ or $z\leq 0$ regions, we overcome this difficulty. The absence of sharp left and right edges of the elongated membrane shown in Fig.~\ref{fig:elongation_config_demo} is a result of this requirement.

\begin{figure}[thb!]
        \includegraphics[width=\linewidth]{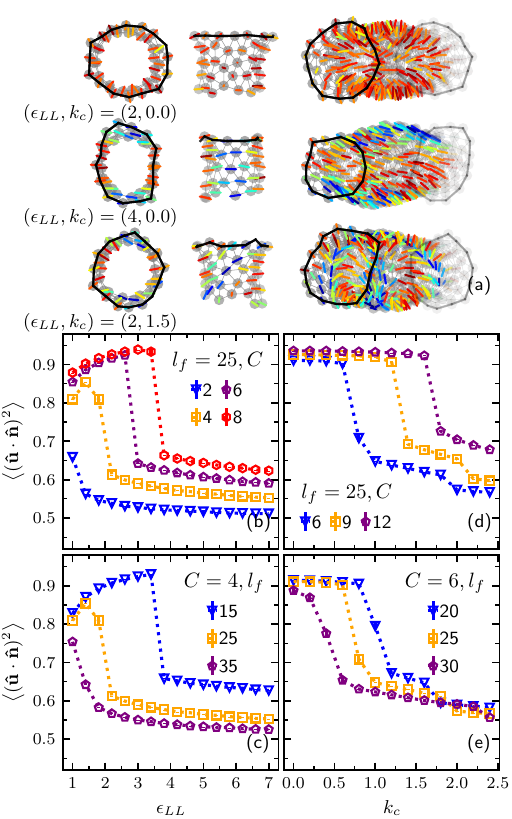}
        \caption{Three phases of the director field (smectic-A, nematic, and cholesteric) in a tubular membrane under force arise from the competition between  the Lebwohl-Lasher constant $\epsilon_{LL}$, tilt coupling $C$, and chirality $k_c$. Here, the number of beads $N=300$, bending modulus $\kappa=50$, and edge tension $\lambda=6$.
        (a) Top view of one end of a membrane of length $l_f=25$ (left column), front view of the same end (middle column, with the beads and bonds in the back sides excluded for better visualization), and oblique view of the tubule (right column). The parameters are for all membrane images are $C=6$, with $(\epsilon_{LL},k_c)=(2,0)$ (top row), $(4,0)$ (middle row), and $(2,1.5)$ (bottom row), respectively. The color coding of the directors is the same as in Fig.~\ref{fig:config_demo}. From top to bottom, the three rows correspond to smectic-A, nematic and cholesteric phases. 
        (b) Average director-normal alignment $\left <(\vu{u}\cdot\vu{n})^2\right >$ versus $\epsilon_{LL}$ for various values of $C$ with $k_c=0$.
        (c) Same as (b) but for various values of $l_f$. 
        (d) $(\vu{u}\cdot\vu{n})^2$ versus chirality $k_c$ for various values of $C$ with $\epsilon_{LL}=2$, 
        (e) Same axes as in (d) but for various values $l_f$.
        The discontinuities in (b) and (c) occur at the smectic-A-nematic transition and in (d) and (e) at the smectic-A-cholesteric transitions.
        }
        \label{fig:smectic_to_walls}
    \end{figure}

    While in the absence of an external force the catenoid membrane exhibits only a cholesteric phase, stretching a cylindrical membrane leads to the appearance of nematic and smectic-A phases in addition to the cholesteric depending on the values of the Lebwohl-Lasher coupling constant $\epsilon_{LL}$, the tilt coupling $C$ and the chirality $k_c$. For sufficiently large $C$ compared to $\epsilon_{LL}$ and with $k_c=0$, the directors align with the local surface normal and form a smectic-A phase as shown in the top row of Fig.~\ref{fig:smectic_to_walls}a. For sufficiently large $\epsilon_{LL}$ compared to $C$ and $k_c=0$, the directors align along a common global direction perpendicular to the $z$ axis and form a nematic phase as shown in middle row of Fig.~\ref{fig:smectic_to_walls}a. Finally, when $k_c$ is nonzero and $C$ is not too large, the directors twist and form a cholesteric phase as shown in the bottom row of Fig.~\ref{fig:smectic_to_walls}a.

    Fig.~\ref{fig:smectic_to_walls}b shows a plot of $\left<(\vu{u}\cdot\vu{n})^2\right>$ versus $\epsilon_{LL}$ for various values of the tilt coupling $C$ and fixed length of the membrane $l_f$. As $C$ increases and $k_c$ remains zero, the critical value of $\epsilon_{LL}$ for the smectic-A-nematic transition increases, indicating that the competition between $C$ and $\epsilon_{LL}$ determines the equilibrium phase. Similarly, the critical value of $k_c$ for the smectic-A-cholesteric transition also increases with increasing $C$ as shown in Fig.~\ref{fig:smectic_to_walls}d, confirming that the competition between $k_c$ and $C$ determines which of these two phases is preferred. On the other hand, Figs.~\ref{fig:smectic_to_walls}c and \ref{fig:smectic_to_walls}e show that increasing $l_f$ \textit{decreases} the critical values of $\epsilon_{LL}$ and $k_c$ at the smectic-A-nematic and smectic-A-cholesteric transitions, respectively. As the length $l_f$ increases, the tubule narrows, increasing the energy penalty for splay of the director field (which is present in the smectic-A phase, see Fig.~\ref{fig:smectic_to_walls}a) and leading to transitions to the nematic and cholesteric phases at lower values of $\epsilon_{LL}$ and $k_c$ respectively.
        
\begin{figure}[!ht]
        \centering
        \includegraphics[width=\linewidth]{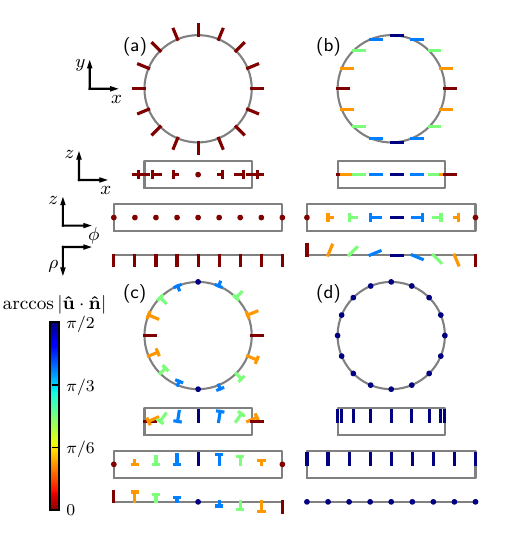}
        \caption{Illustration of perfect 
        (a) smectic-A, (b) and (d) nematic (zero chirality) and (c) cholesteric order on a cylinder whose long axis is the $z$ axis. The directors are indicated by rods and we provide four views of each phase from top to bottom, respectively: looking down the $z$ axis, a view from the side of the cylinder, and two views with the cylinder unwrapped: one looking head-on [i.e., the $(\phi,z)$ plane] and another looking down the $z$ axis. In all figures the director fields have no $z$ dependence. 
        (a) A smectic-A phase where all directors are aligned with the normal to the surface. 
        (b) A nematic phase where all directors point in the $x$ direction. For nonzero but small chirality, the directors will begin to twist. 
        (c) A cholesteric phase where the directors lie in the $(\rho,z)$ plane and rotate about the $\vu{\phi}$ axis. We use a ``nailhead'' representation of the director , where the head of the nail is tilted out of the plane of the figure. (d) A nematic phase with all directors pointing in the z direction.  }
        \label{fig:diagram_walls}
    \end{figure}

    Additional insight into the phases shown in Fig.~\ref{fig:smectic_to_walls} can be obtained from illustrations of perfect nematic, cholesteric and smectic-A ordering on a cylinder (see Fig.~\ref{fig:diagram_walls}). The variation of tilt in our model permits the continuous transformation of the director field between these different phases. The smectic-A phase shown in Fig.~\ref{fig:diagram_walls}a is analogous to a +1 disclination in a planar nematic, and can continuously transform into the nematic phase with directors along $z$ (Fig.~\ref{fig:diagram_walls}d) by escaping into the third dimension.\cite{Meyer1973,cladis-kleman}. Likewise, the cholesteric phase (Fig.~\ref{fig:diagram_walls}c) can transform continuously to the nematic phase with directors along $x$ (Fig.~\ref{fig:diagram_walls}b)  via a rotation of the directors by $\pi/2$ about the radial direction followed by rotations about $z$. Finally, the nematic configurations in Figs.~\ref{fig:diagram_walls}b and~\ref{fig:diagram_walls}d are related by a rotation about $y$. Note that the nematic and cholesteric phases shown in Figs.~\ref{fig:diagram_walls}b and \ref{fig:diagram_walls}c, respectively, have identical tilt energies and this common energy is lower than the tilt energy of the nematic phase shown in Fig.~\ref{fig:diagram_walls}d. In the former two phases there is some alignment of the directors with the local surface normal, whereas in the latter phase all of the directors are perpendicular to the normal.
       
\subsection{Walls}
\label{ssec:Walls}
    We now take a closer look at the director field in the nematic and cholesteric phases of the elongated membrane. In particular, we consider the $\pi$ walls which form as a consequence of the competition between $k_c$, $C$, and $\epsilon_{LL}$. It is helpful to note that $\pi$ walls can appear with two different structures as shown by Helfrich for nematic liquid crystals in a magnetic field.\cite{Helfrich1968} To visualize Helfrich's structures, consider directors with their centers of mass confined to a plane and oriented perpendicular to the plane except in a straight thin domain wall of infinite length. 
    
    If the directors rotate by $180^\circ$ about an axis \textit{perpendicular} to the domain wall, then the wall is a twist $\pi$ wall. Such walls are analogous to Bloch walls in ferromagnets. On a cylindrical surface, twist $\pi$ walls are lines of directors tangent to the surface, with the directors in the wall oriented \textit{parallel} to the wall. An ideal case with twist $\pi$ walls along the $z$ axis is shown in Fig.~\ref{fig:diagram_walls}c. As one crosses a twist $\pi$ wall, the directors rotate by $180^\circ$ about an axis \textit{perpendicular} to the line (i.e., the $\hat{\phi}$ direction in ~\ref{fig:diagram_walls}c). In this example, the $\pi$ walls are not thin because the directors rotate at a uniform rate as the circumference is traversed. 
    
    Returning to the case of molecules confined to a plane, Helfrich noted that $\pi$ walls can also be ``splay-bend" walls with no twist, analogous to N\'{e}el walls in ferromagnets. For directors with centers of mass in a plane, and perpendicular to the surface everywhere but in a thin domain wall, a splay-bend $\pi$ wall has the directors rotating by $180^\circ$ about an axis \textit{parallel} to the wall. Wrapping this plane into a cylinder to make a straight $\pi$ wall analogous to the splay-bend $\pi$ wall, we see that directors in this $\pi$ wall once again lie in the tangent plane of the surface but are oriented $\textit{perpendicular}$ to the wall (see Fig.~\ref{fig:diagram_walls}b). As one crosses this $\pi$ wall, the directors rotate \textit{relative to the surface normal} by $180^\circ$ about an axis \textit{parallel} to the line (the $z$ axis in the figure). Thus, on a cylinder, the analog of the splay-bend $\pi$ wall on a flat surface is a tilt $\pi$ wall. 
    
    More generally, both types of $\pi$ walls need not be parallel to the $z$ axis, and there is not a rigid distinction between the two types of walls. For example, consider a cholesteric state (different from the one shown in~\ref{fig:diagram_walls}c) in which the directors on the cylinder twist about the axis of the cylinder: $\vu{u}=\vu{x}\cos(qz)+\vu{y}\sin(qz)$. This state has two domain walls of mixed type that spiral around the cylinder. As we traverse the domain wall along a circumference, the directors rotate about the $z$ axis relative to the surface normal. Since the domain wall is not along $z$, the axis of rotation makes an angle with the domain wall, indicating it is of mixed type. And if we traverse the domain wall along a path in the surface which is normal to the domain wall, the directors rotate about $z$ relative to a space-fixed axis, again indicating that the wall is partly a twist wall and partly a tilt wall.
             
    \begin{figure*}[htb!]
        \centering
        \includegraphics[width=\linewidth]{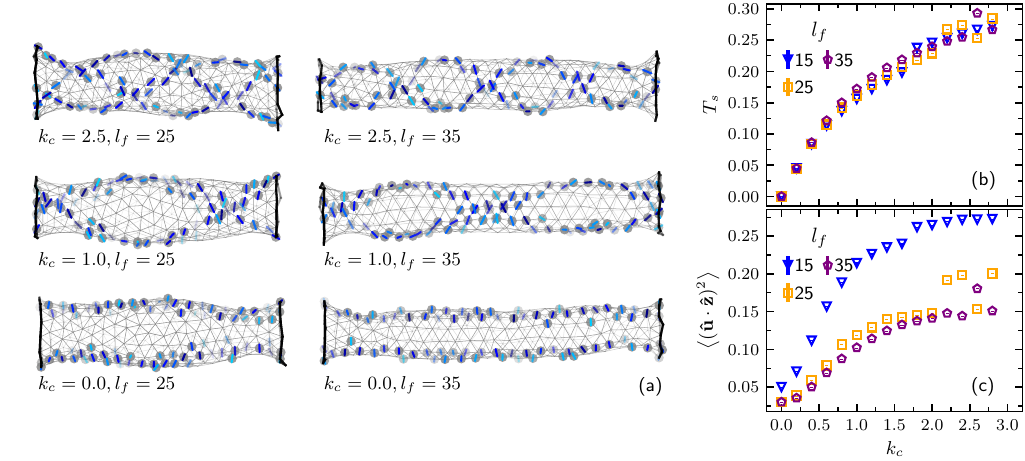}
        \caption{$\pi$ walls in the nematic (zero chirality) and cholesteric phases with number of beads $N=300$, bending modulus $\kappa=50$, edge tension $\lambda=6$, Lebwohl-Lasher constant $\epsilon_{LL}=4$, and tilt coupling $C=4$. 
        (a) Configurations of an elongated membrane with $l_f=25$, 35 (left to right) and $k_c=0$, 1, 2.5 (bottom to top, respectively). Only the triangular mesh and directors with tilt angle $\arccos{|\vu{u}\cdot\vu{n}|}>\pi/3$ are shown. The color bar is the same as in Fig.~\ref{fig:config_demo}. The $\pi$ walls twist around the membrane in a right-handed sense for $k_c>0$. 
        (b) Director field twist $T_s=\left<(\vu{u}_i\cross\vu{u}_j)\cdot \vu{r}_{ij} (\vu{u}_i\cdot\vu{u}_j)\right>_{(i,j)}$, averaged over all bonds $(i,j)$, versus chirality $k_c$ for various values of the membrane length $l_f$. The positive values of $T_s$ indicate that the twist is right-handed. 
        (c) Average projection of the director along the direction of elongation $\left<(\vu{u}\cdot\vu{z})^2 \right>$ versus $k_c$ for different $l_f$. The discontinuities in (b) and (c) correspond to the formation of an additional $\pi$ wall.}
        \label{fig:twisting_wall}
    \end{figure*}
             
    Fig.~\ref{fig:twisting_wall} shows how a change in chirality transforms the tilt walls into twist walls in the cholesteric phase. As shown in Fig.~\ref{fig:twisting_wall}a, the tilt walls start twisting and wrapping about the membrane's cylindrical neck, forming helical shapes as the chirality $k_c$ increases. At the same time, the directors on the wall become parallel to the direction of the wall, as expected for twist walls. These observation are quantified in Figs.~\ref{fig:twisting_wall}b and \ref{fig:twisting_wall}c. The former figure shows the average twist of the director field $T_s=\left<(\vu{u}_i\cross\vu{u}_j)\cdot \vu{r}_{ij} (\vu{u}_i\cdot\vu{u}_j)\right>_{(i,j)}$. Here, the average $\left<\dots\right>_{(i,j)}$ is over all bonds $(i,j)$. The twist increases with the chirality as expected. Fig.~\ref{fig:twisting_wall}c displays $\vu{u}\cdot\vu{z}$, the component of the director along the  elongation direction. This quantity also increases with increasing chirality, indicating that as the directors twist about each other, they also rotate more into the elongation direction and the walls transition from tilt to twist. The discontinuities in  Figs.~\ref{fig:twisting_wall}b and \ref{fig:twisting_wall}c for $l_f=15$ and $l_f=25$ correspond to the formation of an additional $\pi$-wall. This can be seen for the case of $l_f=25$ by comparing the upper left and lower left configurations in Fig.~\ref{fig:twisting_wall}a. Similar discontinuities are not found for the longer membrane with $l_f=35$ whose configuration is shown in the upper right of Fig.~\ref{fig:twisting_wall}a. Due to the smaller diameter of the longer membrane there is insufficient space for an additional $\pi$-wall.

    \begin{figure}[!ht]
        \centering
        \includegraphics[width=1\linewidth]{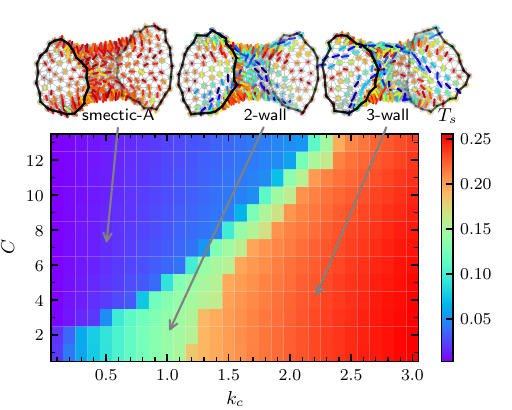}
        \caption{Heat map of the director field twist $T_s=\left<(\vu{u}_i\cross\vu{u}_j)\cdot \vu{r}_{ij} (\vu{u}_i\cdot\vu{u}_j)\right>_{(i,j)}$ for various values of chirality $k_c$ and tilt coupling $C$ with number of beads $N=300$, bending modulus $\kappa=50$, edge tension $\lambda=6$, Lebwohl-Lasher constant $\epsilon_{LL}=2$ for a relative short membrane $l_f=15$. The sharp gradient of the color indicates the transition from smectic-A phase (purple), to 2-wall cholesteric (green) and 3-wall cholesteric (red) phases. Configurations of the membrane are shown at the top with values of $(k_c, C)$ corresponding to points indicated by arrows in the heat map.
        }
        \label{fig:phase_diagram_walls}
    \end{figure}

    Fig.~\ref{fig:phase_diagram_walls} shows the average director field twist $T_s=\left<(\vu{u}_i\cross\vu{u}_j)\cdot \vu{r}_{ij} (\vu{u}_i\cdot\vu{u}_j)\right>_{(i,j)}$ for various values of chirality $k_c$ and tilt coupling $C$. Consistent with the results shown in Fig.~\ref{fig:twisting_wall}, there are three phases: smectic-A, 2-wall (nematic or cholesteric) and 3-wall cholesteric when the membrane is not too thin, the range of chirality $k_c$ of 2-wall phase that separate smectic-A and 3-wall phase get smaller as tilt coupling $C$ increases.
       
    \begin{figure}[htb!]
        \centering
        \includegraphics[width=1\linewidth]{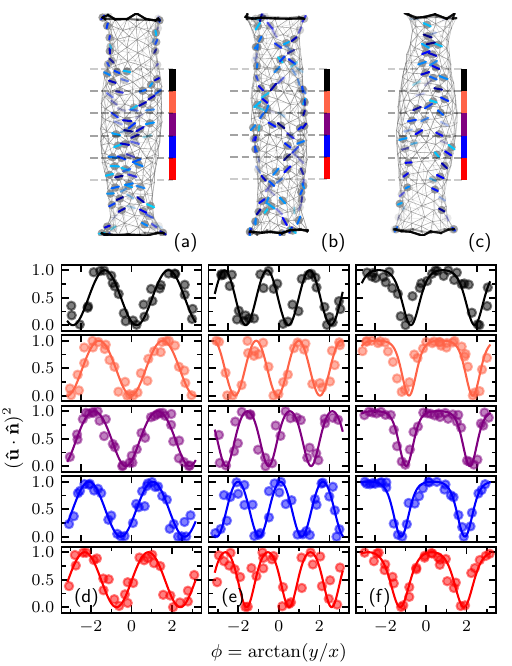}
        \caption{Structure of the $\pi$ walls. Here the number of beads $N=300$, bending modulus $\kappa=50$, edge tension $\lambda=6$, Lebwohl-Lasher constant $\epsilon_{LL}=4$. 
        (a) Configuration of a membrane  with two $\pi$ walls,  tilt coupling $C=2$, length $l_f=25$ and chirality $k_c=0.5$. The horizontal dashed lines and vertical colored lines indicate slices of beads along the $z$ direction. 
        (b) Same as (a) except $k_c=2.5$, and there are now three $\pi$ walls. 
        (c) Same as (a) except $C=10$. 
        (d)-(f) The director-surface normal alignment $(\vu{u}\cdot\vu{n})^2$ as a function of the polar coordinate $\hat{\phi}$ for the configurations shown in (a)-(c), respectively.
        The colors correspond to the slices of the membrane shown in the figure directly above. Each dot corresponds to an individual bead. The curves are fit to the form $(\vu{u}\cdot\vu{n})^2 = \big[\exp(1/\lambda_\phi) - \exp[\sin(m(\phi-\phi_0))/\lambda_\phi]\big]/[\exp(1/\lambda_\phi) - \exp(-1/\lambda_\phi)]$, where $m$ is the number of $\pi$ walls [$m=2$ in (d) and (f) and $m=3$ in (e)], $\lambda_\phi$ in the factor $\exp[\sin(m(\phi-\phi_0))/\lambda_\phi]$ determines the angular width of the $\pi$ walls, and the others terms normalize $(\vu{u}\cdot\vu{n})^2$ to the range $[0,1]$. In (d) and (e), $\lambda_\phi \simeq 3$, yields a good sinusoidal-like fit that indicates that the $\pi$ walls are not sharp; the tilt coupling is relatively small and the directors twist to satisfy their chirality. In (f), the curves are fit with  $\lambda_\phi \simeq 0.5$. The $\pi$ walls here are narrower as the directors tend to align with the surface normal in order to lower the cost of tilt. The location of the $\pi$ walls corresponds to values of $\phi$ where $(\vu{u}\cdot\vu{n})^2=0.$} 
        \label{fig:wall_pitch_analysis}
    \end{figure}
        
    \begin{figure}[htb!]
        \centering
        \includegraphics[width=1\linewidth]{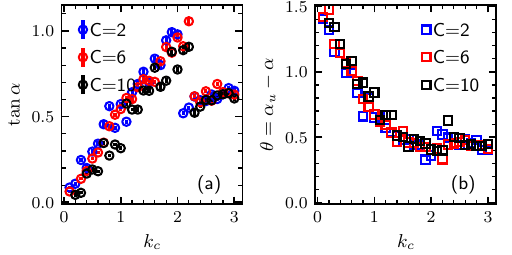}
        \caption{The angle $\alpha$ between the $\pi$ walls and the $z$ axis, and the angle $\theta$ between the walls and directors in the vicinity of the walls. Here the number of beads $N=300$, bending modulus $\kappa=50$, edge tension $\lambda=6$, Lebwohl-Lasher constant $\epsilon_{LL}=4$.
        (a) Plot of $\tan\alpha$ versus chirality $k_c$ for $C=2,6,10$.  The $\tan\alpha$ is obtained from results similar to Figs. \ref{fig:wall_pitch_analysis}d-f using the fit $\phi_0\sim \tan\alpha \left<z\right>/\left<R\right>$, where $\left<z\right>$ is the average $z$ position and $\left<R\right>$ is the average radius of each slice of the nearly cylindrical membrane.
        (b) Plot of $\theta = \alpha_u - \alpha$ versus $k_c$ for tilt coupling $C=2,6$ and 10, As $\theta$ decreases, the directors on and near the wall begin to align with the wall direction and the wall becomes more twist-like in nature. For $k_c\lesssim 2$ two $\pi$ walls are present. The discontinuities in the plots around $k_c=2$ are due to the appearance of a third wall. }
        \label{fig:wall_pitch_result}
    \end{figure}
        
    We further study the structure of the $\pi$ walls by slicing the membrane perpendicular to the $z$ direction, as indicated by the horizontal dashed lines and colored vertical segments in Figs.~\ref{fig:wall_pitch_analysis}a-c. Figs.~\ref{fig:wall_pitch_analysis}a and \ref{fig:wall_pitch_analysis}c show a membrane with two $\pi$ walls for tilt couplings $C = 2,10$, respectively. Figs.~\ref{fig:wall_pitch_analysis}b shows a membrane with three $\pi$ walls with $C=2$, but at a higher value of chirality. Figs.~\ref{fig:wall_pitch_analysis}d-f show $(\vu{u}\cdot\vu{n})^2$, the alignment between the directors and the surface normal in each of the slices shown in Figs.~\ref{fig:wall_pitch_analysis}a-c, plotted against the polar angle $\phi = \arctan(y/x)$, where $(x, y)$ is the location of a bead in the slice. The curves are fit to the form $(\vu{u}\cdot\vu{n})^2 = (e^{1/\lambda_\phi} - e^{\sin{m(\phi-\phi_0)}/\lambda_\phi})/(e^{1/\lambda_\phi} - e^{-1/\lambda_\phi})$, where $m$ is the number of $\pi$ walls ($m=2$ in (d) and (f) and $m=3$ in (e)), $\lambda_\phi$ in the factor $e^{\sin{m(\phi-\phi_0)}/\lambda_\phi}$ determines the angular width of the $\pi$ walls, and the others terms normalize $(\vu{u}\cdot\vu{n})^2$ to the range $[0,1]$. In (d) and (e), $\lambda_\phi \simeq 3$, yields a good sinusoidal-like fit that indicates that the $\pi$ walls are not sharp; the tilt coupling is relatively small and the directors twist to satisfy their chirality. In (f) where the tilt coupling is larger, the curves are fit with  $\lambda_\phi \simeq 0.5$. The $\pi$ walls here are narrower as the directors tend to align with the surface normal in order to lower the cost of tilt. 
         
    Fig.~\ref{fig:wall_pitch_result}b shows the angle $\theta \equiv\alpha_u - \alpha$ as a function of chirality $k_c$, where $\alpha_u$ is the angle between $z$ and directors in the vicinity of the $\pi$ walls which have a tilt angle satisfying $\arccos(|\vu{u}\cdot\vu{n}|)>\pi/3$ and $\alpha$ is obtained from $\tan\alpha$ in Fig.~\ref{fig:wall_pitch_result}a. Thus, $\theta$ measures the orientation of these directors relative to the wall direction, and $\theta$ is zero for twist walls and $\pi/2$ for tilt walls. As $k_c$ increases, $\alpha_u$ decreases, i.e., the directors near the wall rotate towards the $z$ axis. Simultaneously the walls rotate away from the $z$ axis ($\alpha$ increases). The net result is that the angle $\theta$ decreases, appears to plateau around a value of 0.4 radians. The $\pi$ walls thus transition from tilt walls at low values of chirality to more twist-like at higher chirality.

\section{Continuum model for the director field on a cylinder}
\label{ssec:Continuum_model_for_the_director_field on cylinder}
    To better understand the phase behavior seen in the simulations of the previous section and the transformation of the tilt $\pi$ walls at low chirality into  twist walls at higher chirality, and to assess the role of membrane flexibility, we use continuum elasticity theory to study the deformation and director configurations of an infinite nearly cylindrical membrane. The total energy of the membrane is~\cite{kaplan2010theory}
    \begin{equation}
    \begin{split}
        E = &\int \dd A \left\{ \frac{K}{2} \left[(\div{\vu{u}})^2+(\vu{u}\cdot(\curl{\vu{u}})+q)^2 + (\vu{u}\cross(\curl{\vu{u}}))^2\right] \right.
        \\
        & \left.+ \frac{C}{2} \left[1-(\vu{u}\cdot\vu{n})^2\right] +\frac{\kappa}{2}(2H)^2 \right\},
    \end{split}
    \label{eq:E_cal}
    \end{equation}
    where $K$ is the Frank modulus in the one coupling constant approximation, $\vu{u}$ is the unit vector representing the director field, $q$ is the preferred rate of twist, $C$ is the tilt modulus, $\vu{n}$ is the membrane unit normal vector, $\kappa$ is the bending modulus, and $H$ is the mean curvature of the membrane. We have chosen the sign of the chiral term $q$ in eqn~(\ref{eq:E_cal}) to agree with the sign of $k_c$ in our simulation model eqn~(\ref{eq:E_lc}), namely, positive values of $q$ and $k_c$ correspond to a right-handed twist of the director field. Since we study nearly cylindrical shapes, cylindrical coordinates $\phi$ and $z$ are natural. The position of the point $(\phi,z)$ is given by $\mathbf{X}(\phi,z)$, and tangent vectors along the coordinate directions are given by $\mathbf{t}_\mu=\partial_\mu\mathbf{X}$, where $\mu=\phi$ or $z$. Distances along the membrane are determined by the metric tensor $g_{\mu\nu}=\mathbf{t}_\mu\cdot\mathbf{t}_\nu$, with $g$ the determinant of the matrix $g_{\mu\nu}$, $g^{\mu\nu}$ the inverse of the metric tensor, $\dd A=\sqrt{g}\dd \phi \dd z$ the area element, and $\vu{n}=\mathbf{t}_1\times\mathbf{t}_2/\sqrt{g}$ the unit normal. Curvature of the membrane is determined by the curvature tensor $K_{\mu\nu}=\vu{n}\cdot\partial_\mu\mathbf{t}_\nu$, with the mean curvature given by $H=g^{\mu\nu}K_{\mu\nu}/2$. Since the membrane in our continuum model is represented by a vanishingly thin mathematical surface, the gradients in eqn~(\ref{eq:E_cal}) are gradients along the surface of  the membrane: $\boldsymbol{\nabla}=g^{\mu\nu}\mathbf{t}_\mu\partial_\nu$. Thus
    \begin{equation}
        \begin{aligned}
            \div{\vu{u}} &=  g^{\mu\nu} \mathbf{t}_\mu \cdot \partial_\nu\vu{u} \\
            \vu{u}\cdot(\curl{\vu{u}}) &= \vu{u}\cdot(g^{\mu\nu} \mathbf{t}_\mu\times \partial_\nu\vu{u}) \\
            \vu{u}\cross(\curl{\vu{u}}) &= \vu{u}\times(g^{\mu\nu} \mathbf{t}_\mu\times \partial_\nu\vu{u}).
        \end{aligned}
    \end{equation}
    Note that our use of the surface gradient $\mathbf{\nabla}$ in the energy leads to couplings between the directors and the membrane curvature, which is characteristic of membrane models that account for extrinsic (normal) components of derivatives of the director field.\cite{nguyen2013nematic} In the following we denote the angle between the director and the surface normal by $\beta$. 
    
\subsection{Case of large Frank constant}
    We can use the continuum model to explain how the director configuration influences membrane shape in the examples shown in Fig.~\ref{fig:smectic_to_walls}. In the smectic phase, symmetry implies the membrane is a cylinder, with a radius $R$ independent of $\phi$. Observe that the cross section of the membrane in our simulations flattens in the nematic phase (Fig.~\ref{fig:smectic_to_walls}a, left column, middle row), whereas it is nearly circular in the cholesteric phase (Fig.~\ref{fig:smectic_to_walls}a, left column, bottom row).    Apparently, the Lebwohl-Lasher modulus $\epsilon_{LL}$ is large enough compared to $\kappa$ to cause the membrane in the nematic phase to deform so that the director is nearly parallel to the normal over most of the circumference. Therefore, we simplify our theoretical discussion by limiting our analysis to the case that the Frank constant is large compared to the bending stiffness in the continuum model, $K\gg \kappa$, even though $\epsilon_{LL}$ is not large compared to $\kappa$ in our simulation. (Monte Carlo simulations indicate that $K\approx3\epsilon_{LL}$ at low temperature~\cite{CleaverAllen1991}.) In this limit, the only parameter is the dimensionless ratio of the tilt modulus to the bending stiffness, $\chi=CR^2/\kappa$. In the simulations, $\chi\approx1$, but we will see that even when $\chi$ is of order unity, the deflection of the membrane cross-section away from the circular shape is small. Thus, we assume the Frank energy is zero, with $\vu{u}=\vu{x}\cos(qz)+\vu{y}\sin(qz)$, and write the energy to second order in $\chi$ for the deformation $\mathbf{X}=\hat{\boldmath{\rho}}[R+\zeta(\phi,z)]+z\vu{z}$. To enforce the constraint of fixed area, $\int\dd\phi\sqrt{g}=2\pi R$, we introduce a Lagrange multiplier $\eta$ which we expand to first order in $\chi$: $\eta=\eta_0+\chi\eta_1$. Using the same approach as Kaplan et al.~\cite{kaplan2010theory}, we derive the Euler-Lagrange equations. To zeroth order in $\chi$, we find that lea$\eta_0=\kappa/(2R^2)$, which is the tension required to hold a membrane cylinder of radius $R$ in equilibrium.\cite{powers_huber_goldstein2002} Using this result in the Euler-Lagrange equations to first order in $\chi$ yields
    \begin{equation}
        \begin{aligned}
        &R^3\left[\frac{\partial^4\zeta}{\partial z^4}+\frac{2}{R^2}\frac{\partial^2\zeta}{\partial z^2}+\frac{2}{R^2}\frac{\partial^2\zeta}{\partial\phi^2\partial z^2}+\frac{1}{R^4}\frac{\partial^4\zeta}{\partial \phi^4}+\frac{1}{R^4}\zeta\right]\\
        +&\frac{3\chi}{4}\cos(2qz-2\phi)+\frac{\chi}{4}+\frac{\chi\eta_1R^2}{\kappa}=0,
        \end{aligned}
        \label{eqn:ELbigK}
    \end{equation}
    where $\eta_1$ is determined by the constraint of fixed area. Solving eqn~(\ref{eqn:ELbigK}) with the assumption that the minimum radius is at $\phi=0$ and $z=0$ yields
    \begin{equation}
        \frac{\zeta}{R}=-\frac{3\chi}{36+64q^2R^2(2+q^2R^2)}\cos(2qz-2\phi).
    \end{equation}
    In the achiral nematic case, $q=0$, the deformation is similar to that of Fig.~\ref{fig:smectic_to_walls}a, left column and middle row:
    \begin{equation}
        \frac{\zeta}{R}=-\frac{\chi}{12}\cos2\phi.
    \end{equation}
    The extra bending required to make a helical deformation for a chiral membrane greatly reduces the amplitude of the deformation relative to the achiral case. For example, the amplitude of the deformation with $qR=1.5$ is $0.056$ times the amplitude when $q=0$, which is consistent with the fact that the cross-section in the left column and bottom row of fig.~\ref{fig:smectic_to_walls}a is much more circular than the achiral case in the left column and middle row.
 
\subsection{Case of infinite bending stiffness}
    Since the effect of flexibility is small for chiral membranes, we will assume infinite stiffness in the rest of this section and take the membrane shape to be an infinite cylinder of radius $R$ (even in the achiral limit of $q=0$). First we compare the energy of two simple configurations, the smectic-A phase and the cholesteric phase. In the smectic-A phase, the directors point radially outward, $\hat{\mathbf u}_\mathrm{Sm}=\hat{\mathbf{\rho}}$, and the energy per unit length for a cylinder of radius $R$ is $E_\mathrm{Sm}/L=\pi K/R+\pi K q^2R$. In the cholesteric phase, $\hat{\mathbf u}_\mathrm{Chol}=\hat{\mathbf x}\cos qz+\hat{\mathbf y}\sin qz$. Note that this configuration amounts to a thin cylindrical shell of liquid crystal cut from a three-dimensional cholesteric with the pitch axis aligned along the cylinder axis. This configuration has two $\pi$ walls that wind around the surface of the cylinder, and the energy per unit length is $E_\mathrm{Chol}=\pi C R/2$. Comparing these two energies, we see that the cholesteric phase is favored over the smectic phase when $CR^2/K<2(1+q^2R^2)$.

    The directors in the configurations we just considered have no $z$ component, whereas our simulations show that the directors have a nonzero $z$ component when $k_c\neq0$ (Figs.~\ref{fig:twisting_wall}a and~\ref{fig:wall_pitch_analysis}abc). Therefore we construct a ansatz with an $z$ component that allows the directors to rotate toward the $z$ axis:
    \begin{equation}
        \vu{u}(\phi,z) = \frac{(1-\gamma)\vu{u}_1(\phi,z) + \gamma \vu{u}_2(\phi,z)}{|(1-\gamma)\vu{u}_1(\phi,z) + \gamma \vu{u}_2(\phi,z)|},
        \label{eq:u}
    \end{equation}
    where the parameter $\gamma$ ranges from 0 to 1, and  
    \begin{equation}
        \begin{aligned}
            \vu{u}_1(\phi,z) &= \cos\beta(\phi,z) \vu{\rho} - \sin\beta(\phi,z)\vu{\phi} \\
            \vu{u}_2(\phi,z) &= \cos\beta(\phi,z) \vu{\rho} 
            -\sin\beta(\phi,z)(\sin\alpha\vu{\phi}+\cos\alpha\vu{z}).
        \end{aligned}
    \label{eq:un_uc}
    \end{equation}
    The directors make an angle $\beta$ with the surface normal $\vu{\rho}$ with $\beta(\phi,z) = \frac{m}{2}[\phi -(z/R) \tan\alpha]$, $m$ is the number of $\pi$ walls, and $\alpha$ (as defined in the last section) is the angle between the $\pi$ walls and $\vu{z}$. The location of the $\pi$ walls corresponds to $\beta=\pi/2$. 

    \begin{figure}[thb!]
        \includegraphics[width=\linewidth]{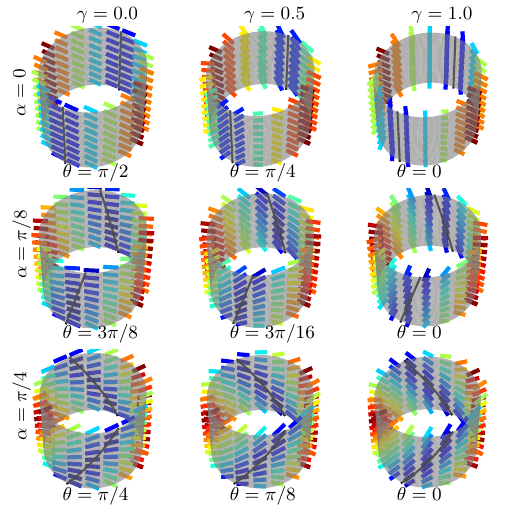}
        \caption{Illustration of the interpolated director field eqns~(\ref{eq:u})-(\ref{eq:un_uc}) with $m=2$ for various values of the interpolation parameter $\gamma$ and tilt angle $\alpha$ of the $\pi$ walls with respect to the $z$ axis. The left column illustrates sample configurations of $ \vu{u}_1(\phi,z)$, depending on the value of $\alpha$, the angle between the $\pi$ walls and the $z$ axis,  while the right column illustrates $ \vu{u}_2(\phi,z)$. Directors are shown as rods and the color bar is the same as Fig.~\ref{fig:config_demo}. The thin black lines near the blue directors indicate the location of $\pi$ walls. The angle $\theta$ is the difference between $\alpha_u$ and $\alpha$, where $\alpha_u$ is the angle between the directors on the wall and the $z$ axis. When $\theta=0$ (right column), the directors on the wall are pointing along the direction of the wall.
        } 
        \label{fig:twomode_config_demo}
    \end{figure}

    Our model is constructed to describe the cholesteric, nematic and smectic-A phases seen in the simulations (Fig.~\ref{fig:smectic_to_walls}a) and in the illustrations of the perfect forms of these phases (Fig.~\ref{fig:diagram_walls}), and to allow for the two types of $\pi$ walls, i.e, tilt and twist. If $m=0$, the director fields $\vu{u}_1$ and $\vu{u}_2$ are equal and describe a smectic-A phase, i.e., $\beta=0$. With $m=2$, $\vu{u}_1$ describes a fully ordered nematic if $\alpha=0$ (Fig.~\ref{fig:diagram_walls}b and the upper left corner of Fig.~\ref{fig:twomode_config_demo}), while for $\alpha \neq 0$, $\vu{u}_1$ describes a cholesteric phase with two $\pi$ walls. If $\alpha$ is close to $\pi/2$, the walls are twist walls, while if $\alpha$ is nearly zero, the walls are tilt walls. This can be seen from the value of the angle $\theta \equiv\alpha_u - \alpha = \pi/2-\alpha$ and recalling that $\theta$ is zero for twist walls and $\pi/2$ for tilt walls. Turning to $\vu{u}_2$, we note that unlike $\vu{u}_1$ it has a $z$ component and the factor $\sin\alpha\vu{\phi}+\cos\alpha\vu{z}$ is a unit vector parallel to a $\pi$ wall. Thus, the director field $\vu{u}_2$ on a wall [$\beta(\phi,z)=\pi/2$] is parallel to the wall direction, and  $\vu{u}_2$ describes a phase with perfect twist walls. Fig. \ref{fig:diagram_walls}c illustrates $\vu{u}_2$ for $m=2, \alpha=0$. There are two $\pi$ walls parallel to the $z$ axis (i.e., perpendicular to the page), one passing through the point at the top and the other through the point at the bottom of the figure. We see from the figure that the directors rotate by $180^\circ$ as either wall is traversed, with a rotation axis parallel to $\hat{\phi}$. Because our ansatz is the smectic-A state when $m=0$, and is like an interpolation between the nematic state and a cholesteric state with two twist $\pi$ walls only when $m=2$, we restrict $m$ to the values 0 and 2.
        
    Fig.~\ref{fig:twomode_config_demo} shows sample configurations of the $m=2$ director field for additional values of $\alpha$ and $\gamma$. In summary, if $m=0$ the model exhibits a smectic-A phase, and if $m=2$ it exhibits a nematic phase if $\alpha= 0$ or a cholesteric phase otherwise. The cholesteric phase can have either tilt or twist walls depending on the values of $\alpha$ and $\gamma$.
    
    \begin{figure}[thb!]
        \centering
        \includegraphics[width=\linewidth]{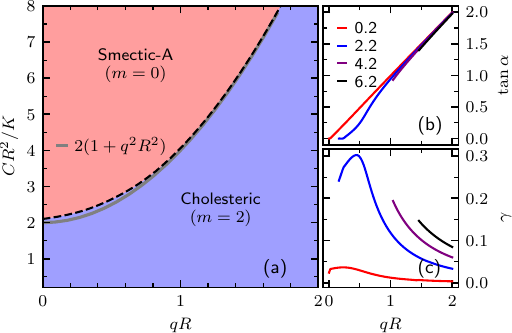}
        \caption{Phase diagram obtained from the interpolated director field eqn (\ref{eq:u}) with $m=0$ and $2$. (a) Phase diagram obtained from the normalized energy difference $\Delta E' R^2/C = (E'_{m=2}-E'_{m=0})R^2/C$ for various values of $qR$ and $CR^2/K$. The dashed line is a continuous transition from the smectic-A to the cholesteric phase. At $qR=0$ the cholesteric becomes a nematic phase up until the critical value of $CR^2/K$ where it transitions to the smectic-A phase. (b) $\tan\alpha$ of the $m=2$ director field (at the minimum value of energy) versus $qR$. The curves are labeled by the values of $CR^2/K$.  and begin at the critical value of $qR$ for the $m=0$ to $m=2$ transition shown in (a). (c) Similar to (b), the value of  the interpolation parameter $\gamma$ of the $m=2$ director field versus $qR$ at the energy minimum. }
        \label{fig:two_mod_del_E}
    \end{figure}
    
    We numerically minimize the energy eqn~\eqref{eq:E_cal}, as a function of $\alpha$ and $\gamma$ for fixed values of $qR$, $CR^2/K$, and $m$. Then, we compute the normalized energy difference $\Delta E' R^2/C = (E'_{m=2}-E'_{m=0})R^2/C$ and find the phase diagram shown in Fig.~\ref{fig:two_mod_del_E}a. Similar to the results from our Monte Carlo simulations in Fig.~\ref{fig:phase_diagram_walls}, the critical $qR$ for the transition from the smectic-A phase to the nematic ($q=0$) or cholesteric phase increases with increasing $CR^2/K$. As shown in Fig.~\ref{fig:two_mod_del_E}b, $\tan\alpha$ also increases with chirality $qR$, indicating a twisting of the $\pi$ walls as chirality increases. Fig.~\ref{fig:two_mod_del_E}c shows $\gamma$ as a function of $qR$ and indicates that the system accommodates the increase of chirality by transitioning from nematic (pure $ \vu{u}_1$ with $\alpha = 0$) to cholesteric ($m=2$) order. When $C$ is very small (0.2), we see from Fig.~\ref{fig:two_mod_del_E}c that the cholesteric phase is described by a nearly pure $ \vu{u}_1$ director field which is illustrated in the left column of Fig.~\ref{fig:twomode_config_demo}. Fig.~\ref{fig:two_mod_del_E}b indicates that $\alpha$ grows and $\theta$ decreases as $qR$ increases (see Fig.~\ref{fig:twomode_config_demo}). As chirality increases, the $\pi$ walls tilt, the directors on the walls rotate towards the direction of the wall, and the tilt walls transition to having a more twist-like character, as we saw in the simulations (see Fig.~\ref{fig:wall_pitch_result}c). The curves shown in Figs.~\ref{fig:two_mod_del_E}b and \ref{fig:two_mod_del_E}c begin at the critical value of $qR$ for the smectic-A to cholesteric transition shown in Fig.~\ref{fig:two_mod_del_E}a. The peak in  Fig.~\ref{fig:two_mod_del_E}c for $C=2.2$ appears because of the flatness of the transition line. Similar peaks for the larger values of $C$ do not appear because they would correspond to points in the smectic-A phase of Fig.~\ref{fig:two_mod_del_E}a. 
    
    We note that the value of $\gamma$ remains less than 0.3 and is generally less than 0.1. Thus, the director field is nearly  $\vu{u}_1$ with a small $\vu{u}_2$ component, which is why the simple balance $E_\mathrm{Sm}=E_\mathrm{Chol}$ mentioned at the beginning of this subsection gives a good approximation to the phase boundary shown in Fig.~\ref{fig:two_mod_del_E}a, and is also why the minimizing value of $\tan\alpha$ is so close to $qR$ ( Fig.~\ref{fig:two_mod_del_E}b). These results are consistent with our simulations (Fig.~\ref{fig:wall_pitch_result}c) which showed that the $\pi$ walls, while no longer purely tilt in character, do not become pure twist walls as chirality increases.

    The tilt energy appearing in both eqns~(\ref{eq:E_lc}) and (\ref{eq:E_cal}) has the same mathematical form as the interaction of directors with a magnetic field imagined to be normal to the surface of the cylinder. For sufficiently large field, the cholesteric twist will be unwound and the system will become a nematic. This problem was studied in a flat geometry by de Gennes\cite{DEGENNEScholesteric} and Meyer\cite{Meyer_cholesteric} who found a critical field proportional to chirality. In our case of a cylindrical geometry, the cholesteric twist can unwind either through the formation of a nematic phase (zero chirality) or a smectic-A phase for sufficiently large $C$. The curvature of the surface makes the nematic state of complete alignment different from the smectic-A state of complete alignment with the ``external field'' of the surface normal. Thus, the transition line between the cholesteric and smectic-A phase does not go through the origin of the phase diagram shown in  Fig.~\ref{fig:two_mod_del_E}a.
    
\section{Conclusion}
\label{sec:conclusion}
    In this paper, we further developed a general model \cite{ding2021deformation} of chiral membranes with orientational order and edges and carried out simulations of membranes with multiple edges. We found that membranes can form disks, catenoids and trinoids as the magnitude of chirality increases and when the number of edges allows. The formation of catenoids and trinoids is accompanied by the appearance of a cholesteric phase where $\pi$-walls wrap around the membrane and connect different edges. For the two-edge membranes, pulling on the opposite edges makes the membrane thinner and leads to a cylindrical shape. The directors on the elongated membrane can form additional phases besides the cholesteric seen in the force-free case. When there is no chirality, the directors can either align with the surface normal and form a smectic-A phase or form a nematic phase with all directors pointing along a single global direction, depending on the strength of the tilt coupling. Once chirality is nonzero, a cholesteric phase appears for sufficiently low tilt coupling. At low chirality, the $\pi$ walls are of the tilt variety. As chirality increases, the walls transform to the twist variety common to the cholesteric phase.
    
    Our model provides a general framework for simulating not only colloidal membrane made of chiral filaments but also the general problem of liquid crystals on deformable surfaces. The current formulation of our model has some limitations. First, the model is restricted to membranes made of a single components, and many shapes including high-order saddle, catenoid and handles emerge when the membranes are made of mixtures of filaments of different lengths.\cite{saddles} A natural extension of the current model to account for mixtures would be to use moduli in the energy that vary with position across the membrane. Second, the current model does not allow topological changes of the triangular mesh. Thus, the edges are only able to shrink to a small triangle instead of fully disappearing. A future development of the model could overcome this limitation with the implementation of an edge removal and creation update that allows a change in the number of edges during the simulation and would also allow a nucleation of a hole in the initial configuration.

\section*{Conflicts of interest}
    There are no conflicts to declare.
    
\section*{Acknowledgements}
    We are grateful to Timothy Atherton, Federico Cao, Zvonimir Dogic, Chaitanya Joshi, and Zifei Liu for helpful conversations.
    This work was supported by the National Science Foundation through Grant No. CMMI-2020098.



\balance


\bibliography{rsc} 
\bibliographystyle{rsc} 

\end{document}